\documentclass[twocolumn,secnumarabic,amssymb,amsmath,aps,prb,floatfix]{revtex4-1}
\usepackage{graphicx}
\usepackage{dcolumn}
\usepackage{bm}
\usepackage{amsmath}
\usepackage{braket}
\usepackage{array}
\usepackage{txfonts}

\begin{document}

\title{Crystal Field Excitations in Breathing 
Pyrochlore Antiferromagnet Ba$_3$Yb$_2$Zn$_5$O$_{11}$}
\author{T. Haku$^1$, M. Soda$^1$, M. Sera$^2$, 
K. Kimura$^2$, S. Itoh$^3$, T. Yokoo$^3$, 
T. Masuda$^1$}
\affiliation{
$^1$ {\it Institute for Solid State Physics, The University of Tokyo, Kashiwa, Chiba 277-8581, Japan}
\\
$^2$ {\it Division of Materials Physics, Graduate School of Engineering Science, Osaka University, Toyonaka, Osaka 560-8531, Japan}
\\
$^3$ {\it Institute of Materials Structure Science, High Energy Accelerator Research Organization, Tsukuba, Ibaraki 305-0801, Japan}
}

\begin{abstract}
Inelastic neutron scattering measurement is performed on a 
breathing pyrochlore antiferromagnet $\rm Ba_3Yb_2Zn_5O_{11}$.
The observed dispersionless excitations are 
explained by a crystalline electric field (CEF) 
Hamiltonian of Kramers ion Yb$^{3+}$ 
of which the local symmetry exhibits $C_{3v}$ 
point group symmetry. 
The magnetic susceptibility previously reported is 
consistently reproduced by the energy scheme of 
the CEF excitations. 
The obtained wave functions of the ground state Kramers doublet 
exhibit the planer-type anisotropy. 
The result demonstrates that $\rm Ba_3Yb_2Zn_5O_{11}$ 
is an experimental realization of 
breathing pyrochlore antiferromagnet 
with a pseudospin $S=1/2$ having easy-plane anisotropy.
\end{abstract}

\maketitle

\section{INTRODUCTION}
Geometrical frustration in magnetic materials disturbs the
development of long-range order and induces novel 
states at low temperatures.\cite{Geo_Rev,Balents}
A three-dimensional network of corner-sharing tetrahedra, 
i.e., the pyrochlore lattice, 
is one of the most interesting systems. 
The geometrical configuration prohibits the arrangement of 
the spins that satisfy the lowest energy of all spin bonds, 
leading to the ground state sensitive to perturbations 
including single-ion anisotropy, two-ion anisotropy, 
and lattice distortion.
Indeed the variety of the magnetic phases have been  
reported in 
rare-earth pyrochlore compounds in $R_2$Ti$_2$O$_7$.\cite{PYRO::Rev} 
In case of $R$ = Dy and Ho 
spin-ice states emerge\cite{EL::R2Ti2O7} 
due to the geometrical frustration
induced by ferromagnetic interaction and the Ising 
anisotropy.\cite{RTi2O7::1,RTi2O7::2,RTi2O7::3,RTi2O7::4} 
In contrast in $\rm Yb_2Ti_2O_7$ the easy-plane type anisotropy of 
Yb ion enhances quantum effect, leading to quantum 
spin liquid state.\cite{XXZ} 

In a series of pyrochlore titanates $R_2$Ti$_2$O$_7$ 
rare earth ions carry magnetic moment. 
The $J$ multiplets of the ions are lifted by crystalline 
electric field (CEF) from octahedral ligands 
with the point symmetry $D_{3d}$. 
In case of $R$ = Dy, Ho, and Yb, the ground state 
is doublet and the first excited energies are larger than 
20 meV,\cite{EL::R2Ti2O7} meaning that the degree of freedom of the 
magnetic moments are approximately two in the temperature 
range of $T << 200$ K. 
The magnetic moments are, thus, 
regarded as pseudospins $S = 1/2$ 
with anisotropies determined by the wave functions 
of the ground state. 
In most cases the localized orbitals of $f$-electrons 
give small magnetic interactions, 
and, therefore, rare-earth magnets can be interacting 
spin systems at low temperatures. 
Since the spin anisotropy is one of key features for 
the emerged magnetic state, experimental study on CEF 
is crucial particularly for the initial stage of 
the research on rare-earth magnets. 

\begin{figure}[htbp]
\begin{center}
	\includegraphics[width=86mm]{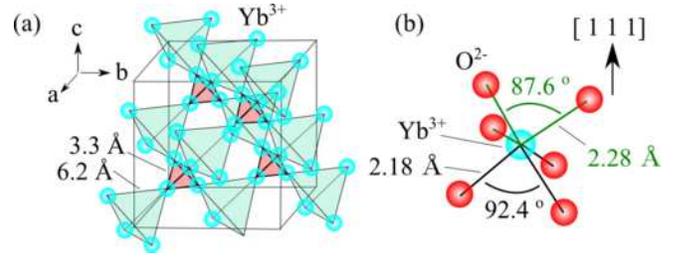}
	\caption{\label{fig::cry}
		Crystal structure of $\rm Ba_3Yb_2Zn_5O_{11}$.
		(a) Breathing pyrochlore network of $\rm Yb^{3+}$ ions.
		(b) Local structure of $\rm YbO_6$, which has 
$C_{3v}$ symmetry.
		The symmetry axis is [1 1 1].
	}
\end{center}
\end{figure}

{\it Breathing pyrochlore} lattice, 
i.e., alternating arrays of corner-sharing large and small tetrahedra, 
is interesting in terms of realization of the perturbative expansion 
method in theoretical calculation.\cite{pyro::theo::2} 
The original model compounds were reported in 
3{\it d}-transition metal spinels $\rm LiInCr_4O_8$
and $\rm LiGaCr_4O_8$.\cite{br::pyro::1}
Recently rare-earth-based compound was found in 
$\rm Ba_3Yb_2Zn_5O_{11}$ where Kramers ion
$\rm Yb^{3+}$ carries pseudospin $S = 1/2$.\cite{br::pyro::2}
Fig. \ref{fig::cry}(a) shows the breathing pyrochlore network of 
$\rm Yb^{3+}$ ions.
Distances between $\rm Yb^{3+}$ ions
of small and large tetrahedra are 3.3 \AA\ and 6.2 \AA,
respectively.
The space group is $F\overline{4}3m$ with cubic symmetry.
The local structure of $\rm YbO_6$ is a distorted 
octahedron with the $C_{3v}$ point group symmetry 
as shown in Fig. \ref{fig::cry} (b).

No phase transition were observed 
in the specific heat and magnetic
measurement in $0.4~{\rm K} < T < 300~{\rm K}$.
A temperature dependence of the magnetic susceptibility 
was reasonably reproduced using 
cubic symmetry for the CEF Hamiltonian down to 30 K, 
even though $\rm YbO_6$ originally exhibited
lower symmetry. 
This indicated that local distortion of a YbO$_6$ octahedron 
is small enough to 
be approximated as a structure with cubic symmetry.
It has been reported that the ground state of CEF Hamiltonian
is a Kramers doublet, and that the first excited state is 
a quartet state having an excitation energy of $E = 45.1 {\rm meV}$
\cite{br::pyro::2}.
Up to room temperature, the excitation energy is sufficiently high
so that $\rm Yb^{3+}$ moment is regarded as 
an isotropic pseudospin $S = 1/2$. 
The bulk magnetic properties in the low temperatures 
were explained by the presumed isotropic Heisenberg 
$S=1/2$ model having non-magnetic ground state. 
For the further understanding of the ground state, however, 
the information of the spin anisotropy is 
important and, therefore, the investigation on the CEF  
is indispensable. 

In the present study, we performed inelastic neutron (INS)
scattering measurement to identify the precise CEF Hamiltonian.
Three excitations were observed at
$\hbar\omega =38.2, 55.0$ and $68.3\ {\rm meV}$,
and all were qualitatively consistent with the energy 
spectrum of the $C_{3v}$ point
group of the local $\rm YbO_6$ structure, 
consisting of four Kramers doublets.
The energy of the first excited state is consistent with that
estimated by the magnetic susceptibility 
measurement, such that $\rm Yb^{3+}$
ions are regarded as the ion having 
pseudospin $S = 1/2$ at low temperatures. 
The $C_{3v}$ symmetry leads to the anisotropic ground state 
wave functions with the spin anisotropy of $q = 3.31$ 
for $J_{xy} = qS_{xy}$ and $p= 2.35$ for $J_z = pS_z$, 
where ${\bm J}$ is total angular momentum and 
${\bm S}$ is effective spin $S = 1/2$ operator.  

\section{EXPERIMENTAL DETAILS}
The inelastic neutron scattering measurement (INS) was performed 
by using high resolution chopper spectrometer (HRC) installed 
in J-PARC/MLF. 
We used 17.7 g powder sample synthesized by a solid state reaction 
method.
The sample was set in an Al-can filled with exchange He gas.
Measurements were performed at
$T = 3$, $200$ and $300 {\rm \ K}$
using an $\rm ^4He$-type closed-cycle refrigerator. 
The $T_0$ chopper for the elimination of 
fast neutrons was set at 50 Hz and a collimator of 
1.5$^{\circ}$ was installed in front of sample. 
The initial neutrons were monochromated by 
type ``S'' Fermi chopper with the frequency of 600 
Hz to obtain the neutron energy of $E_i$ = 154.4 meV with 
the instrumental resolution (full width at half maximum, FHWM)
of 5.5 meV at the elastic position. 
Preliminary INS measurement with $E_i$ = 150 meV was performed at MARI 
specrometer installed in ISIS.

\section{RESULTS}

\begin{figure}[tb]
\begin{center}
	\includegraphics[width=90mm]{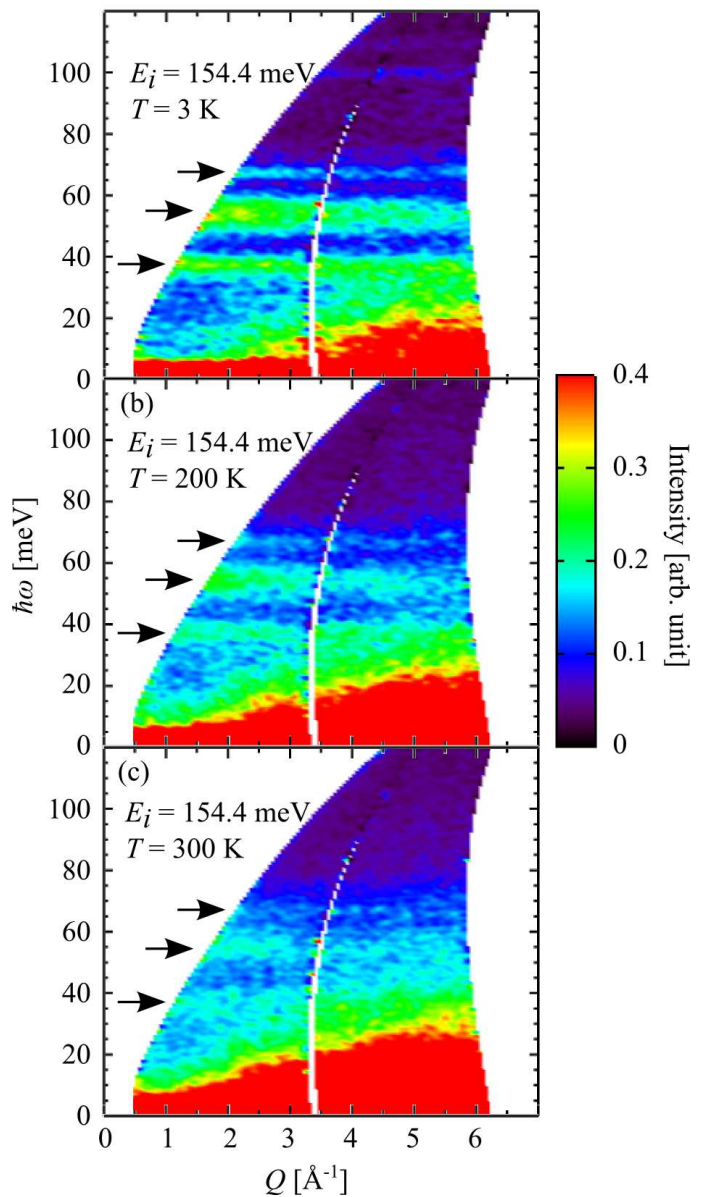}
	\caption{\label{fig::CEF::HRC::map}
		Contour plots of inelastic neutron spectra measured at
		at 
		(a)$T = 3$,
		(b)$T = 200$, and
		(c)$T = 300{\ \rm K}$.
		The arrows indicate dispersionless excitations
		at $\hbar \omega =$ 
		$38.2$, $55.0$ and $68.3 {\ \rm meV}$.
	}
\end{center}
\end{figure}

Figure \ref{fig::CEF::HRC::map}(a), (b), and (c) show 
contour maps of the INS spectrum
at $T = 3$, $200$, and $300 {\rm\ K}$, respectively.
The arrows indicate three dispersionless excitations 
at $\hbar \omega =38.2, 55.0$ and $68.3 {\ \rm meV}$. 
The intensities of these excitations 
decrease with the increase of $Q$, and 
they decrease with the increase of the temperature. 
None of the excitation-energy values
changes in the observed temperature range.
These behaviors are consistent with that of excitations of 
a magnetic cluster. 
At $\hbar \omega \lesssim 30 {\rm meV}$ we observed broad excitation
of which the intensity increases 
with the $Q$, and 
it increases with the temperature. 
The excitation is, thus, regarded as phonons. 
At $\hbar \omega \sim 100$ meV flat stripe is 
observed in Fig. \ref{fig::CEF::HRC::map}(a). 
Since no intensity was observed at 100 meV in preliminary measurement at MARI (not shown), 
the stripe is regarded as artifact. 

\begin{figure}[tb]
	\includegraphics[width=85mm]{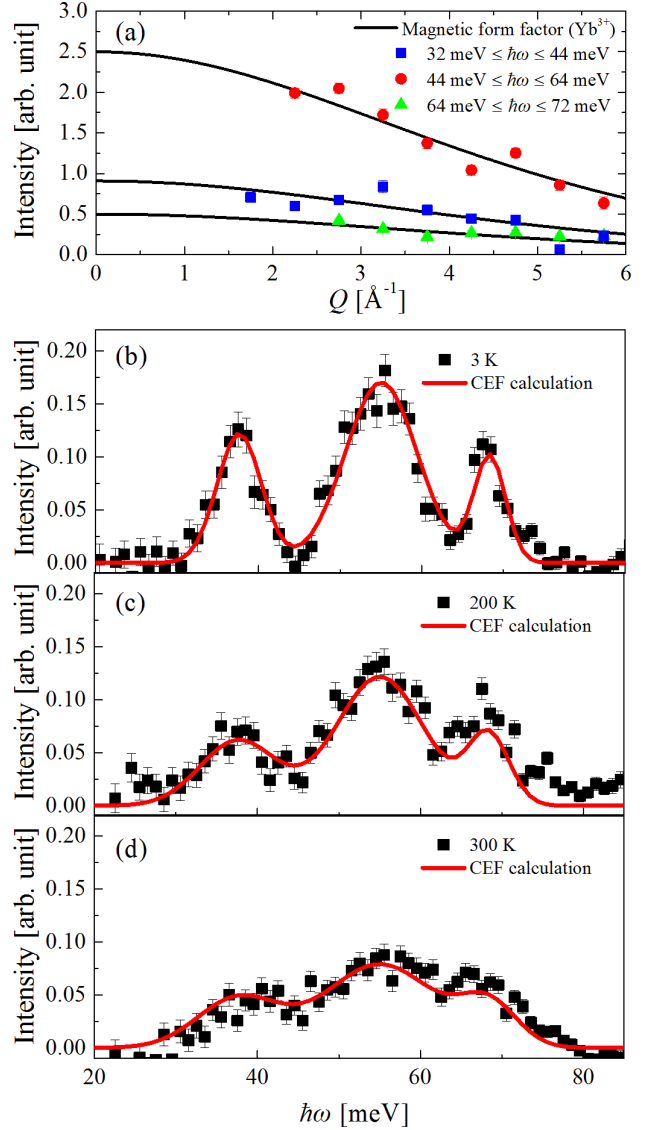}
	\caption{\label{fig::CEF::Qdep}
		(a) $Q$ dependence of neutron intensities for three modes.
		The symbols are the data obtained 
                by integrating the intensities in the 
		ranges of
		$32 {\rm meV} \le \hbar\omega \le 44 {\rm meV}$,
		$44 {\rm meV} \le \hbar\omega \le 64 {\rm meV}$ and
		$64 {\rm meV} \le \hbar\omega \le 72 {\rm meV}$
		at $T = 3\ {\ \rm K}$.
		The solid curves show the $Q$ dependence of the
		magnetic form factor of $\rm Yb^{3+}$
		in Ref. \onlinecite{mff}.
		The symbols of (b), (c) and (d)
		show the $\hbar \omega$ dependences 
                of the neutron intensities 
                at $T =$ 3, 200 and 300 K.
		The intensities are integrated in the range of 
		$2.8 {\rm \AA}^{-1} \le Q \le 3.2{\rm \AA}^{-1}$.
		The red curves show the calculations (see the text).
			}
\end{figure}

Figure \ref{fig::CEF::Qdep}(a) shows the $Q$ dependences of 
the intensities of the 
three modes. 
The data are obtained by integrating the intensities in the 
ranges of $32~{\rm meV} \le \hbar \omega \le 44~{\rm meV}$, 
$44~{\rm meV} \le \hbar \omega \le 64~{\rm meV}$, and 
$64~{\rm meV} \le \hbar \omega \le 72~{\rm meV}$. 
The intensities decrease with the increase of $Q$, and they 
follow the magnetic form factor of the $\rm Yb^{3+}$ ion
indicated by the solid curves.\cite{mff}  
These excitations, thus, 
derive from the CEF of the $\rm Yb^{3+}$ ion.

\begin{figure}[tb]
\begin{center}
	\includegraphics[width=86mm]{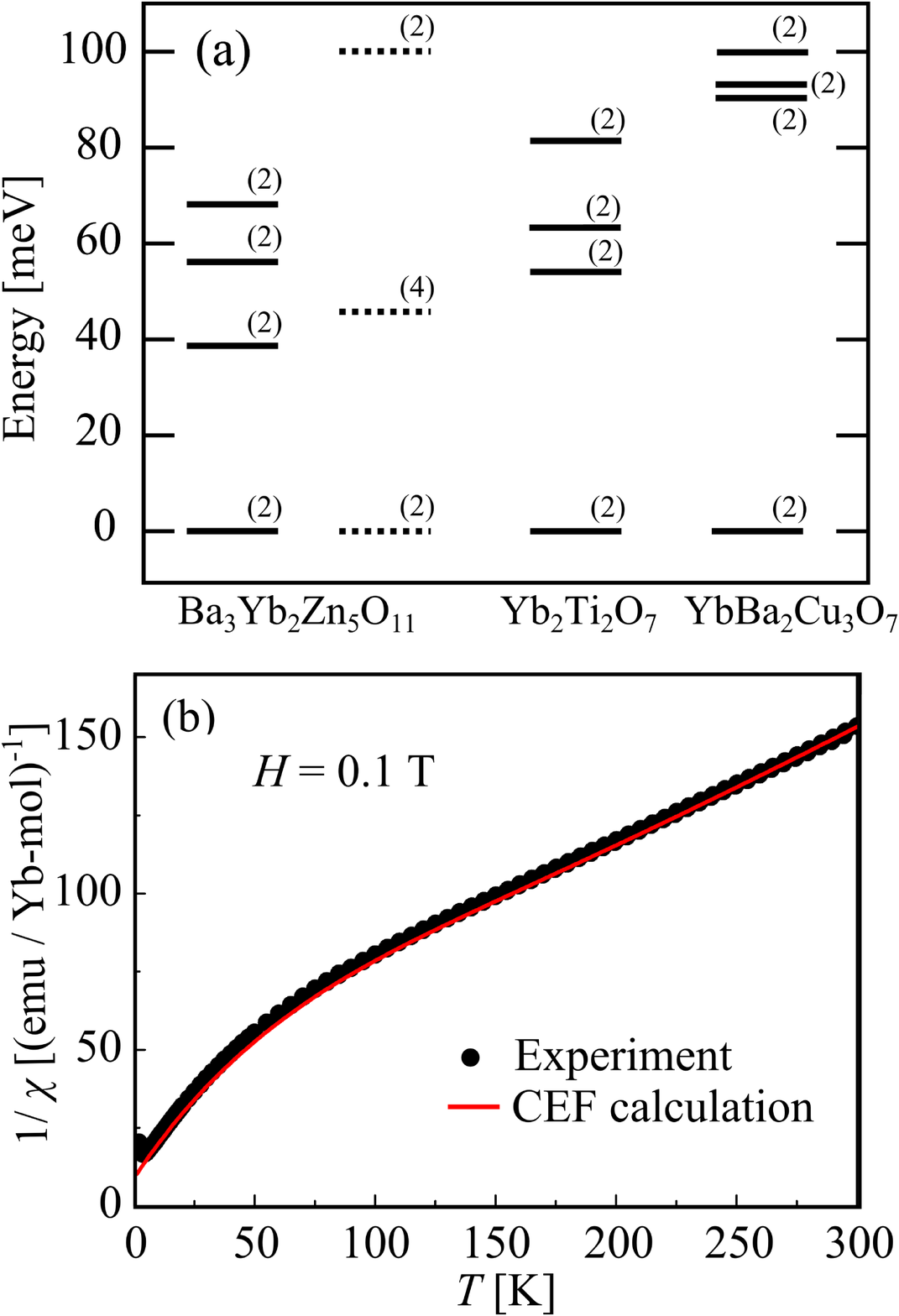}
	\caption{\label{fig::CEF::el}
		(a)
		Energy diagrams of $\rm Yb^{3+}$ compounds,
		$\rm Ba_3Yb_2Zn_5O_{11}$
		\cite{br::pyro::2},
		$\rm Yb_2Ti_2O_7$ 
		\cite{EL::R2Ti2O7}
		and
		$\rm YbBa_2Cu_3O_7$
		\cite{YbBCO}.
		The solid and dotted lines indicate
		the levels obtained by neutron scattering studies
		and bulk measurement, respectively.
		(b)
		Magnetic susceptibility curve at $H$ = 0.1 T.
		The solid circles are the experimental values cited in ref. 16.\ 
		The solid curve is a calculated value using
		the parameter set from the first line of Table \ref{CEF_p} 
		and an antiferromagnetic molecular field constant $\lambda = -9 \ {\rm emu/mol}$.
	}
\end{center}
\end{figure}

The symbols shown in Fig. \ref{fig::CEF::Qdep}(b), \ref{fig::CEF::Qdep}(c),
and \ref{fig::CEF::Qdep}(d) indicate 
$\hbar \omega$ dependence of the integrated intensities 
at $T$ = 3, 200, and 300 K, respectively, 
where the integral range is 2.8 {\AA}$^{-1} \le Q \le 3.2 {\rm \AA}^{-1}$. 
The phonon contributions have been subtracted 
by assuming that the intensity
is proportional to $Q^2$ and also to the Bose factor. 
The intensities of the three modes decrease with the increase of 
the temperature. 
Since 
the occupancy of the ground state decreases with the increase of 
the temperature and the occupancy is one of coefficients in magnetic neutron 
cross section, 
all of the excitations derive from 
the transitions between 
the ground state and excited states. 
This means that 
the Yb$^{3+}$ ion is the four level system as described by solid bars in 
the first row in Fig. \ref{fig::CEF::el}(a). 
The result is in contrast with the three level system 
assumed in the previous study\cite{br::pyro::2} as described by 
dotted bars in the second row. 
The first excited energy of the present study is, however, close to 
that of previous one. 
This means that the assumption of pseudospin $S = 1/2$ 
for the Yb$^{3+}$ ion\cite{br::pyro::2} is good in the low temperatures.

The three peaks at $T$ = 3 K in Fig. \ref{fig::CEF::Qdep}(b) are 
fit by Gaussian functions, and the integrated intensities 
will be used to compare the calculation in the next section. 
The estimated FWHMs at 
$\hbar \omega = $ $38.2$, $55.0,$ and $68.3$ meV
are $\Delta\hbar\omega =$
$5.5$, $8.9$ and $3.9\ {\rm meV}$, respectively.
The FWHM at $55.0$ meV is wider than the instrumental energy resolution.
This indicates that there is either a lowering of 
symmetry in the local structure 
of $\rm YbO_6$ or a coupling of CEF and phonon excitations.
In addition, at $T = 200$ and $300$ K, 
the peaks widthes bcome broadened. 
This indicates that the relaxation of the CEF excitation is 
enhanced thorough the coupling with lattice. \cite{Lovesey00} 

\begin{table*}[htbp]
	\begin{center}
		\caption{
			 Values for $B^m_n$ of $\rm Ba_3Yb_2Zn_5O_{11}$ 
obtained by INS experiment in the present study and those by the magnetic 
susceptibility measurement in the previous study.\cite{br::pyro::2} 
			}
		\label{CEF_p}
		\begin{tabular*}{175mm}{p{35mm}p{19mm}p{19mm}p{19mm}p{19mm}p{19mm}p{19mm}}
		\hline
		\hline
		& $B^0_2$ & $B^0_4 \times 10^{2}$ & $B^3_4 \times 10^{2}$ 
		& $B^0_6 \times 10^{4}$ & $B^3_6 \times 10^{4}$ & $B^6_6 \times 10^{4}$  \\
		\hline
		Present study (meV)
			& $0.0737$		 & $-2.76$
			& $-33.6$		 & $ 6.10$
			& $ 8.70$ & $ 88.2$ \\
		Previous study (meV)
			& $0$			 & $ 3.45$
			& $-97.4$		 & $ 3.06$
			& $ 37.9$ & $ 29.5$ \\
		\hline
		\hline
		\end{tabular*}
	\end{center}
\end{table*}
\section{DISCUSSIONS}

The obtained energy diagrams of $\rm Ba_3Yb_2Zn_5O_{11}$ 
is different from that of simple cubic symmetry for the local 
structure around $\rm Yb^{3+}$ ion 
assumed in a previous study.\cite{br::pyro::2}
Instead the diagram is similar to the case of oxides
of which the local structure is distorted, 
$\rm Yb_2Ti_2O_7$ or
$\rm YbBa_2Cu_3O_7$, where 
the first excited states with fourfold degeneracy in the cubic symmetry 
are lifted into a pair of doublets. 
We, hence, analyze the CEF excitation based on the precise
ligand symmetry, $C_{3v}$, 
in $\rm Ba_3Yb_2Zn_5O_{11}$.

The CEF Hamiltonian is as follows,
\begin{align}
{\mathcal H}_{CEF} = B_2^0O_2^0 + B_4^0 O_4^0 +  B_4^3 O_4^3 
\nonumber
\\
+ B_6^0 O_6^0 +  B_6^3 O_6^3 + B_6^6 O_6^6
,
\label{eq::H_CEF}
\end{align}
where $B^n_m$ and $O^n_m$ are CEF parameters and Steven's operators, 
respectively.
\cite{CEF::Hami::B,CEF::Hami::O}
The neutron cross section of the CEF excitations is,
according to the dipole approximation,\cite{CEF::sqw}
\begin{align}
&I_{calc} (\bm{\kappa},\hbar\omega) = 
r_0^2 \frac{k'}{k}
\sum _{\alpha, \beta=x,y,z}
(\delta_{\alpha \beta}- \hat{\kappa}_ \alpha \hat{\kappa}_ \beta) 
F^2({\kappa}) 
\nonumber \\
&\times p_\lambda
\braket{\lambda|J_\alpha|\lambda'}
 \braket{\lambda'|J_\beta|\lambda}
 \delta \left( \hbar \omega - E_{\lambda'} + E_{\lambda} \right).
\label{INS::CEF}
\end{align}
Here ${\bm \kappa}$ is scattering vector, 
and $\hat{\bm \kappa}$ is the normalized one. 
The initial and final states of the CEF excitations are denoted by 
$\ket{\lambda}$ and $\ket{\lambda'}$.
The probability of state $\ket{\lambda}$ is represented as $p_\lambda$.
$F(\kappa)$ is the magnetic form factor.\cite{mff}
The value of $r_0$ is $-0.54 \times 10^{-15}\ {\rm m}$.
Powder averaged intensity 
$I_{\rm calc}^{\rm powder} ({\kappa},\hbar\omega)$ 
is obtained 
in order to compare the calculation to the experimental data. 
The integrated intensities for three modes 
at $T$ = 3 K obtained in Fig. \ref{fig::CEF::Qdep}(b) 
are fit by 
$I_{\rm calc}^{\rm powder} ({\kappa},\hbar\omega)$. 
The solid curves are 
$I_{\rm calc}^{\rm powder} ({\kappa},\hbar\omega)$ 
convoluted by Gaussian functions having FWHMs obtained in 
previous section. 
The fit to the data is reasonable. 
The obtained CEF
parameters are summarized in Table \ref{CEF_p}.
The wave functions of the ground states are 
\begin{align}
\ket{\pm} = \mp0.537 \ket{\pm \frac{7}{2}}
	- 0.805 \ket{\pm \frac{1}{2}}
	\pm 0.251 \ket{\mp \frac{5}{2}}.
\end{align}
The solid curves in 
Figs. \ref{fig::CEF::Qdep}(c) and \ref{fig::CEF::Qdep}(d) 
are the calculation using the obtained CEF parameters. 
Since the peaks are broadened with the increase of the temperature, 
we take the FWHMs for the peaks as fitting parameters. 
The calculation reasonably reproduces the data. 
When a trigonal axis of the $\rm YbO_6$ structure is chosen as a quantization axis
and a pseudospin operator $\bm S$ is defined as
$J_x = q S_x$,
$J_y = q S_y$,
$J_z = p S_z$,
$p$ and $q$ are as follows, 
\begin{align}
\pm \frac{1}{2} p &= \braket{\pm|J_z|\pm}
= \pm \frac{1}{2} \times 2.35 \label{eq4}
\\
q &= \braket{\pm|J_\pm|\mp}
=3.31. \label{eq5}
\end{align}
The result means that a wave function of the ground 
state has planar-type anisotropy. 

The magnetic susceptibility $\chi$ is calculated 
as follows.\cite{CEF::chi}
\begin{align}
\chi = \chi_{dia} + \chi_{CEF} / (1 - \lambda \chi_{CEF})
\label{eq::H_CEF::chi}
\end{align}
Here $\chi_{CEF}$ is the magnetic susceptibility determined 
by the CEF Hamiltonian.
$\chi _{dia}$ is the diamagnetic susceptibility 
fixed to be $-4 \times 10^{-4}$emu/mol-Yb.
$\lambda$ is a molecular field constant defined as
${\bm H}_{eff} = {\bm H} + \lambda {\bm M}$, where 
$\bm H$, $\bm M$ and ${\bm H}_{eff}$ are an external 
field, magnetization and an effective  
molecular field, respectively.
The solid curve in Fig. \ref{fig::CEF::el} (b)
is the calculated one with the parameters obtained by the present INS 
experiment 
and an antiferromagnetic molecular 
field constant $\lambda = -9 \ {\rm mol/emu}$.
The experimental data 
are from Ref. \onlinecite{br::pyro::2}.
The calculation reasonably reproduced the experimental data.
The CEF parameter obtained by INS experiment 
reproduces the magnetic susceptibility measured independently.

\section{CONCLUSION}

We performed a neutron scattering experiment 
on a powder sample of $\rm Ba_3Yb_2Zn_5O_{11}$ 
in order to identify the CEF Hamiltonian. 
The neutron spectrum was explained by 
four Kramers doublets of Yb$^{3+}$ 
ion. 
The obtained CEF parameters revealed that the Yb$^{3+}$ 
ions are regarded as the pseudospin $S$ = 1/2 
having easy-plane anisotropy. 
Next challenge is further investigation on the low-energy 
spin dynamics of the breathing pyrochlore spin system by using 
a cold neutron spectrometer. 

\section*{Acknowledgements}
T. Haku was supported by the Japan Society for the Promotion of Science 
through the Program for Leading Graduate Schools (MERIT).
This work was supported by JSPS KAKENHI Grant in Aid for 
Scientific Research (B) Grant No. 24340077.
The neutron scattering experiment at HRC spectrometer in KEK 
was approved by the Neutron Scattering
Program Advisory Committee of IMSS, KEK (Proposal No. 2013S01 and 2014S01) 
and ISSP.
Preliminary neutron experiment was performed at MARI spectrometer in 
ISIS, Rutherford Appleton Laboratory. 
Dr. J. Taylor is greatly appreciated for his experimental support at MARI. 
Travel expenses for the experiment performed using 
MARI at ISIS, UK, were supported by General User Program for 
Neutron Scattering Experiments, Institute for Solid State Physics, 
The University of Tokyo (proposal no. 14522), at JRR-3, 
Japan Atomic Energy Agency, Tokai, Japan.

\end{document}